\documentclass[12pt]{article}

\usepackage{graphics}
\usepackage{amssymb}
\usepackage{cite}
\usepackage{epsfig}
\usepackage{psfrag}

\def\eq#1{(\ref{#1})}
\def\Eq#1{Eq.~(\ref{#1})}

\def\beq{\begin{equation}}
\def\eeq{\end{equation}}
\def\beqa{\begin{eqnarray}}
\def\eeqa{\end{eqnarray}}
\def\bet{\begin{tabular}}
\def\eet{\end{tabular}}
\def\del{\partial}
\newcommand{\ex}[1]{{\rm e}^{#1}} \def\ii{{\rm i}}
\newcommand{\Tr}{{\rm Tr}}

\renewcommand{\a}{\alpha}

\newcommand{\e}{\epsilon}
\newcommand{\ve}{{\vec{\e}}}

\newcommand{\bxi}{\mbox{\boldmath $\xi$}}

\renewcommand{\Im}{{\rm Im}\,}

\begin{document}


\setcounter{page}{1}

\begin{flushright}

{DFTT 25/2005}\\

\end{flushright}

\vspace{.2cm}

\begin{center}
{\Large \bf Charged open strings in a background field and
  Euler-Heisenberg effective action\footnote{Partially supported by the European Commission,
under RTN program MRTN-CT-2004-0051004 and by the Italian MIUR under
the contract PRIN 2003023852.}.} \\

\vskip .2cm

{\bf  Stefano Sciuto}\\
{\sl Dipartimento di Fisica Teorica, Universit\`a di Torino}\\
 and {\sl  INFN, Sezione di Torino},
{\sl Via P. Giuria 1, I-10125 Torino, Italy}\\

\vskip .1cm

\vskip .2cm

\end{center}

\begin{abstract}
This talk is based on work made with L. Magnea and R. Russo ~\cite{noi}.
We give an explicit expression of the multiloop partition function of
open bosonic string theory in the presence of a constant gauge field
strength.
The Schottky parametrization allows to perform  the field
theory limit, which at two-loop level  reproduces the 
 Euler-Heisenberg effective action for adjoint scalars minimally
 coupled to the background gauge field.

\end{abstract}
\vskip .2cm

String theory achievements are very impressive: maybe the most
important is that the consistency
between quantum mechanics and special  relativity in the description
of the motion of a string implies the presence of gravitons (and of
their supersimmetric partners) and therefore yields a consistent
quantum theory of (super)gravity.
But the richness (and the problems) of string theory as a fundamental theory
cannot be  discussed now; here we will take a much more
modest attitude just looking to string theory as a natural and very
fruitful embedding for Quantum Field Theory; open strings, in
particular, represent a
natural generalization of charged particles since they couple, through
their endpoints, to  gauge fields.
A peculiar feature of this embedding is {\bf open-closed string
  duality}: exchanging the role of the worldsheet coordinates $ \tau$
and $\sigma$, an annulus spanned by the time evolution of an open
string can also be seen as a cylinder described by free propagation
of a closed string. 
This property has far reaching consequences as it relates gravity
(described by closed strings) with Yang Mills theories (open strings);
here we will use it only as a technical device for computing open
string amplitudes not otherwise accessible.

We will focus on the case of open bosonic strings and 
will study the multiloop partition function in the presence of a
constant Yang-Mills field strength $F$. We will then use the string
formula in the low-energy limit to recover the Euler-Heisenberg~
\cite{Heisenberg:1935qt}
effective action for a gauge field coupled to adjoint scalars; we have
performed the calculation at one and two loops, but in principle it
could be done at any perturbative order.
 From the mathematical point of view, the open string
diagram is represented as usual by a Riemann surface with $g + 1$
boundaries; however the presence of  external fields $F$ introduces
twisted boundary conditions along some of the boundaries. As a
consequence, the basic geometric building blocks of the string
amplitude, such as the determinant of the Laplace operator on the
Riemann surface, are deformed by $F$. The necessary ingredients to
derive the multiloop partition function for charged open strings were
assembled in~\cite{Russo:2003tt,Russo:2003yk}, developing earlier
studies~\cite{Verlinde:1986kw,Alvarez-Gaume:1987vm}.
  We will now outline the derivation of the partition function for {\it
charged} open strings attached to $g+1$ (stacks of N) D-branes, {\it i.e.} open
  strings with mixed boundary conditions
$\left[ \del_\sigma X^i + \ii \,\del_\tau X^j 
F_j^{~i \,(A)} \right]_{\sigma = 0} = 0~$,
where $F^A$ is the constant gauge field strength on the $A$th D-brane
($A = 0, \ldots, g$).

A direct computation of the charged partition function in the open
string channel is difficult, mainly because  for $F\neq 0$ the string 
coordinates
have a non-integer mode expansion. However the open-closed string
duality under the exchange $\tau \leftrightarrow - \sigma$  allows  
to go to
the closed string channel by using boundary states satisfying $
\left[ \del_\tau X^i - \ii \,\del_\sigma X^j 
F_j^{~i \,(A)} \right]_{\tau = 0} | B \rangle_{F_A}  =  0~.$
The field $F$ can always be put in a block-diagonal form, so for the sake of 
simplicity we will take the space-time indices to be in the plane $i,j
= 1,2$ and we will write $F_{12}^{(A)} = - F_{21}^{(A)} = \tan(\pi
\epsilon^A)$.
The computation of vacuum diagrams can be done along the lines 
of~\cite{Frau:1997mq};
to be specific, we take the external boundary to have
Neumann boundary conditions ($F^{(0)} = 0$), so that $\vec{\e}$ is a
vector with $g$ components, denoted by $\e_\mu$, encoding the values
of the gauge field on the remaining $g$ boundaries.
 The result is
\beq
Z_F^c (g) = \left( \prod_{\mu = 1}^g \frac{1}{\cos \pi \e_\mu} \right) 
\int \left[ d Z \right]_g^c \, {\cal R}_g \left(q_\a, \vec{\e}\right)~,
\label{clch}
\eeq
where\footnote{We refer to the Appendices 
of~\cite{Russo:2003tt} for a short explanation of the Schottky
parametrization used in \eq{mis}; here we just recall that for a Riemann
surface of genus $g$ the Schottky group is freely generated by $g$
projective transformations $S_\mu$, each of them characterized by one
multiplier $q_\mu$ and two fixed points $\eta_\mu$ and $\xi_\mu$; for
closed string surfaces related by a modular transformation to
open string world surfaces one has  $\xi_\mu=  \bar\eta_\mu$.}
 
\beq
\left[d Z \right]_g^c = 
\frac{1}{d V_{abc}} \prod_{\mu = 1}^{g} \left[
\frac{d q_\mu \, d^2 \eta^c_\mu \, (1 - q_\mu)^2}{q_\mu^2 \, 
(\eta^c_\mu - \bar\eta^c_\mu)^2} \right]
{\prod_{\a}}' \left(\prod_{n = 1}^\infty (1 - q^n_\a)^{-d}
\prod_{n = 2}^\infty (1 - q^n_\a)^2 \right) \, ,
\label{mis}
\eeq
represents the $F = 0$ result\footnote{$d V_{abc}$  signals that we
  have to fix three real
variable among the $\eta$'s to take into account the overall
projective invariance; the superscript $c$  reminds that the parameters
describe closed string exchanges among the various
boundaries; ${\prod_{\a}}'$ means the product over all the primitive
elements $T_\a$ of the Schottky group ({\it i.e.} those which can not be
written as powers of other elements); finally $d$ is the
dimensionality of 
space-time.}, while
the $\vec{\e}$ dependence is encoded in the factor
\beq
{\cal R}_g \left(q_\a, \vec{\e}\right) =  \frac{{\prod_\alpha}' 
\prod_{n = 1}^\infty (1 - q_\alpha^n)^2}{ {\prod_\alpha}' 
\prod_{n = 1}^\infty \left( 1 - \ex{ - 2 \pi \ii \vec{\e} \cdot \vec{N}_\a} 
q^n_\alpha \right) \left( 1 - \ex{ 2 \pi \ii \vec{\e} \cdot \vec{N}_\a} 
q^n_\alpha \right)}~.
\label{ratioclosed}
\eeq
Here  the $\mu^{\rm th}$ entry of $\vec{N}_\a$ counts how many times
the Schottky generator $S_\mu$ enters
in the element of the Schottky group $T_\a$, whose multiplier is
$q_\a$.
The factors of $1/\cos(\pi \e)$ in \Eq{clch},
 are nothing but a rewriting of the Born-Infeld contribution
to the boundary state normalization (see for
instance~\cite{DiVecchia:1999fx}).

\Eq{clch} contains all the information about the interaction among
charged open strings, but  is written in the closed string
representation; as it stands, its  low energy limit  would
describe scattering of gravitons.
To get instead Yang Mills theory, we must go back to the open string
channel; this can be achieved by  performing 
the modular 
transformation $\tau^c = -\tau^{-1}$, where  $\tau^c$ and  $\tau$ are
the period matrices in the closed and open channel.
To do this, at first, we rewrite the
products over the Schottky group in terms of geometrical objects with
simple transformation properties under the modular group, like
$\theta$ functions, differentials and the prime form; then, we perform
the modular transformation; as a last step, we go back to the Schottky
parametrization, which is the most appropriate for performing the low
energy limit. The technical tool needed in this derivation is the
higher-genus generalization of the Jacobi formulae expressing $\theta$
functions as products. These formulae can be derived by exploiting
bosonization identities in two dimensions in presence of the
twists $\vec{\e}$
\cite{Russo:2003tt,Russo:2003yk}. 
The final result for the string effective action in the open string channel is
\beq
Z_F (g) =  \left(\frac{\ex{ 2 \pi \ii \e_g} - 1}{\prod_{\mu = 1}^g 
\cos{\pi \e_\mu}} \right) 
\int \left[d Z \right]_g \, \left[ \ex{- \ii \pi \vec{\e} \cdot \tau \cdot
\vec{\e}} \; \frac{\det \left(\tau \right)}{\det \left(\tau_\ve \right)}
\; {\cal R}_g \left(k_\a, \vec{\e} \cdot \tau \right) \right]\,,
\label{effstr}
\eeq
where 
\beqa
 \left[d Z \right]_g \ & = &  \frac{1}{d V_{a b c}} \prod_{\mu = 1}^g \left[
d k_\mu d \eta_\mu d \xi_\mu 
\frac{(1 - k_\mu)^2}{k_\mu^2 (\xi_\mu - 
\eta_\mu)^2} \right] 
\left[\det \left(\Im \tau \right) \right]^{-\frac{d}{2}} \nonumber \\
&& \times \, {\prod_\alpha}' \left[\frac{
\prod_{n = 2}^\infty (1 - k_\alpha^n)^2}
{\prod_{n = 1}^\infty (1 - k_\alpha^n)^{d} }\right]~
\label{scg}
\eeqa
is the open string channel transcription of eq.\eq{mis}; the crucial
new object in eq.\eq{effstr} is $\tau_\ve$, which is built by means of
 $g-1$ Prym differentials (generalization of the
usual $g$ abelian differentials to the case of {\it twisted} boundary
conditions) and which reduces to the usual  period matrix $\tau$ for
$\ve=0$; the explicit expression of $\tau_\ve$ in terms of the Schottky 
parametrization can be found in  \cite{noi}.

Now we want to show that the low energy limit of this string
configuration reproduces a theory of adjoint scalars coupled to a
background Yang Mills field.
To be more precise  let us consider a classical background
$U(N)$ gauge field ${\cal A}_\mu$, represented as a hermitean $N
\times N$ matrix ${\cal A}_\mu = \sum_a A_\mu^a \, T_a$, where $T_a$
($a = 0, \ldots, N^2 - 1$) are $U(N)$ generators, coupled to a quantum
massive scalar field, also in the adjoint
representation, $\Phi = \sum_a \varphi^a \, T_a$.
The gauge field configuration corresponding to a single charged brane
 is a diagonal ${\cal A}_\mu$ matrix, with all
eigenvalues vanishing except one, 
$\left({\cal A}_\mu \right)_{A B} = A_\mu \, \delta_{A,N}
\delta_{B,N}$.  This
choice of background breaks the symmetry in color space, so that the
matter ``multiplet'' $\Phi$ will have both neutral ($\sigma$ and $\Pi$)  and
charged ($\bxi$ and $\bxi^\dagger$) components
with respect to ${\cal A}_\mu$. The
lagrangian is given by
\beqa
{\cal L} & = & \Tr \left[ \partial_\mu  \Pi \, \partial^\mu \Pi 
\right] + \frac{1}{2} \partial_\mu \sigma \, \partial^\mu \sigma +
D_\mu \bxi^\dagger D^\mu \bxi - m^2 \, \Tr \left( \Pi^2 \right) - \frac{1}{2}
m^2 \sigma^2 \nonumber \\ & - & m^2 \bxi^\dagger \bxi + \frac{2}{3} \lambda
\, \Tr \left( \Pi^3 \right) + \sqrt{2} \, \frac{\lambda}{6} \sigma^3 + 
\frac{\lambda}{\sqrt{2}} \, \sigma \bxi^\dagger \bxi + \lambda
\, \bxi^\dagger \Pi \, \bxi~,
\label{lag2}
\eeqa
where $\Pi$ is a hermitean $(N - 1) \times (N - 1)$ matrix
representing a field in the adjoint representation of $U(N - 1)$,
$\bxi$ is a complex vector in the fundamental representation of $U(N -
1)$, while $\sigma$ is a singlet real field, and the abelian covariant 
derivative is defined by $
D_\mu \bxi = \partial_\mu \bxi + {\rm i} A_\mu \bxi~$.

 Now we perform the low-energy limit of the string partition
function in \Eq{effstr}
sending   the typical length of the string $\sqrt{\a'}$  to zero
and taking into account that both the moduli describing the shape of the
Riemann surface and the physical magnetic field $B$ (which must be
kept fixed) are dimensionful. The
logarithms of the multipliers of Schottky transformations, for
example, are associated with the length of the corresponding loops by
setting $\log k_\mu = - T_\mu/\a'$, where $T_\mu$ is the sum of the
Schwinger parameters associated with the propagators forming the
$\mu^{\rm th}$ loop. Moreover, taking only one boundary 
charged and setting $\e_g \equiv \e \neq 0$, the field theory limit
is defined by $\tan(\pi \e) = 2 \pi \a' B$, which implies $\e = 2
\a' B + {\cal O} \left(\a'^3 \right)$. Other dimensionful quantities
are an overall normalization constant and the scalar self-coupling
$\lambda$, 
which must be matched  with the string coupling $g_S$.
 In order to isolate the contribution of charged scalars circulating 
in the loops we have to look to the powers of the
multiplier in a Taylor expansion of the integrand for small
$k_\mu$. For the bosonic string, this expansion starts with
$k_\mu^{-2}$, a sign of the tachyonic instability; this singularity
can however be readily regularized by recalling that the tachyon mass
squared is $m^2 = -1/\a'$ and setting~\cite{DiVecchia:1996kf} $
d k_\mu/k_\mu^2 = -\exp (T_\mu/\a') d T_\mu/\a' =
 -\exp ( - m^2 T_\mu) d T_\mu/\a'$.

We skip the easiest one loop case, where  \Eq{effstr} reproduces the
results of Ref.~\cite{Bachas:1992bh}, for the magnetic case.
 At two loops, we will take
advantage of the projective invariance by setting 
$\eta_1 = 0$, $\xi_1 = \infty$ and $\xi_2 = 1$.
Looking only for the 1PI vacuum graphs, we set  \cite{Frizzo:1999zx}  
 $ k_1 = \exp ( - (t_1 + t_3)/\a')~, \quad
k_2 = \exp ( - (t_2 + t_3)/{\a'} )~, \quad
\eta_2 = \exp( - t_3/{\a'})~, $ 
where $t_i$ are the Schwinger parameters associated with the three
propagators in the diagram ($t_2$ and $t_3$ to the loop with a charged
boundary and $t_1$ to the neutral particle).
After the expansion in powers of $k_\mu$ of the 
infinite series and products over the Schottky group, one finally gets
the effective action in the low energy limit:
\beq
W_{\xi \Pi}^{(2)} (m, B) \, = \, - {\rm i} \, V_d \, 
\frac{\lambda^2}{(4 \pi)^d} \,
\frac{(N - 1)^2}{4} \int_0^\infty d t_1 d t_2 d t_3
\, {\rm e}^{- m^2 (t_1 + t_2 + t_3)} \, \Delta_0^{- \frac{d}{2} + 1} \,
\Delta_B^{-1} \, ,
\label{twores}
\eeq
where $
\Delta_B = \sinh (B t_2) \sinh (B t_3)/B^2 + t_1
\sinh \left[ B \left(t_2 + t_3 \right) \right]/B~$
and  $\Delta_0 = \lim_{B \to 0} \Delta_B = t_1 t_2 + t_1 t_3 + t_2
t_3$, for the diagram with the exchange of the neutral particle $\Pi$;
the only change when the exchanged particle is $\sigma$ is in the
 color factor, with the replacement $(N - 1)^2 \to N - 1$.
One can check that this result, obtained by the low energy limit of
the vacuum string amplitude with twisted boundary conditions, exactly
agrees with the field theoretical result that one can compute
starting from the lagrangian \eq{lag2}.

The procedure could be generalized to derive the 
Euler-Heisenberg effective action for pure Yang-Mills theory; the
starting point is always~\Eq{effstr}, but one has to isolate the
contributions to the loop integrals of the first
excited state in the spectrum of the open bosonic string, which is a
massless vector. In practice, this means that the expansion in
multipliers of the various geometrical objects appearing in
\Eq{effstr} has to be pushed one order higher. Finally, the extension to 
superstrings would allow to study also the coupling with 
charged fermions.

\providecommand{\href}[2]{#2}\begingroup\raggedright\endgroup

\end{document}